\documentclass{aa}
\usepackage{graphicx}
\begin{document}
   \title{Further Study of the Gamma-Ray Bursts Duration 
Distribution }


   \author{I. Horv\'ath   
          }


   \institute{ Dept. of Physics,
              Bolyai Military University, Budapest,
              POB 12, H-1456, Hungary\\
Present address:        
              NASA, Goddard Space Flight Center, Greenbelt, MD 20771 \\
              \email{hoi@bjkmf.hu}                  
                             }

   \date{Received , 2002; accepted }

   \abstract{ Two classes of gamma-ray bursts have been identified so far, 
characterized by 
durations shorter and longer than approximately 2 seconds. 
In 1998 two independent papers indicated the existence of the third 
class of the bursts roughly duration between 2 and 10 seconds. 
In this paper, using the full BATSE Catalog, the maximum likelihood
estimation is presented, which gives a 0.5\% probability to having only 
two subclasses.
The Monte-Carlo simulation confirms this probability, too. 
   
   \keywords{ Gamma rays: bursts, theory, observations
 --  Methods: data analysis, observational, statistical
                   }
   }

   \maketitle


\section{INTRODUCTION}

In the BATSE Current Catalog (\cite{M}) there are 2702 Gamma-Ray 
Bursts (GRBs), of which 2041 have duration information. 
\cite{K1} have identified two types of GRB based on durations, 
for which the value of $T_{90}$ (the time during which 90\% of the 
fluence is accumulated) is smaller or larger than 2 s, respectively, and
exhibits an acceptable bimodal log-normal ("two-Gaussian") fit.
This bimodal
distribution has been further quantified in another paper (\cite{K}),
where a two-Gaussian fit was made, however the best parameters of the fit
were published in \cite{mcb} and \cite{Kos}.

Previously we have published an article (\cite{HO}), where both 
two- and three-Gaussian
fits were made using the $\chi^2$ method, which gave 
a 99.98\% significance the third Gaussian is needed.
This is an agreement with the result of \cite{MUK}, who used a
multivariate analysis and found that the probability of existence of two 
groups, rather than three ones, is less than $10^{-4}$.
\cite{HA} also confirmed this result by statistical
clustering analysis. However, they suggested
that the third group was caused by instrumental biases
(\cite{HAK1}, \cite{HAK2}).
Recently \cite{br} have applied automatic classifier algorithms
and obtain three different classes of GRBs.
Add also that the intermediate subgroup shows a remarkable angular 
distribution on the sky (\cite{bal98}, \cite{bal99}, 
 \cite{me00b}, \cite{li01}).

Although the high probabilities for the occurrence of third 
(intermediate in duration)
subgroup are suggestive, the existence of the third subgroup is 
a matter of debate.
Hence further studies concerning this subclass are highly needed.
In order to make further progress
in quantifying this classification, one of the issues which needs 
to be addressed is an evaluation of the probabilities associated 
with the bimodal, or - in general - with the multimodal distribution.
In this paper we take another
attempt at the trimodal log-normal distribution of $T_{90}$, evaluating a new small
probability of the assumption that the third subclass is a chance 
occurrence. 

   \begin{figure}
   \centering
\resizebox{\hsize}{!}
{\includegraphics[angle=-90,width=7cm]{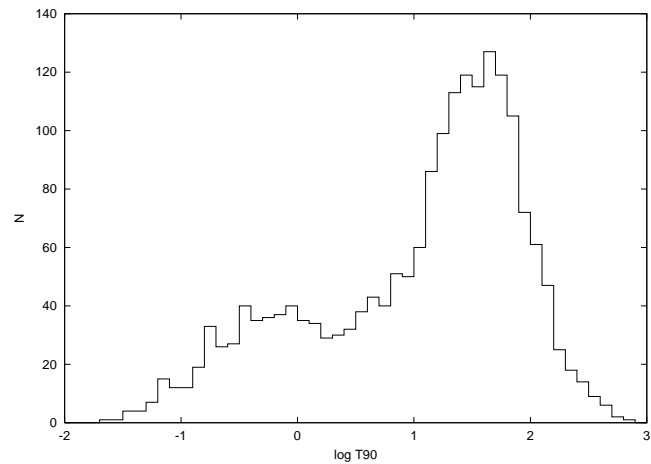}}
      \caption{Duration distribution of the observed BATSE bursts.  }
   \end{figure}

In Section 2  bi and a tri-modal log-normal fits have been made using
the maximum likelihood method. In Section 3 a hundred Monte-Carlo
simulations have been taken confirm the low probability.
In Section 4 a triggering systematic effect has been discussed.
Finally the conclusions are given in Section 5.

\section{FITS IN $\log T_{90}$ }

For this investigation we have used a smaller set of 1929 burst 
durations in 
the Current Catalog, because only they have peak flux information as well.
We use the $T_{90}$ measures provided in this data set.
Figure 1. shows the distribution of $\log T_{90}$.

   \begin{table}
      \caption[]{Two Gaussian fit of 
the GRBs duration distribution}
         \label{KapSou}
     $$ 
         \begin{array}{cccc}
            \hline
            \noalign{\smallskip}
     & Duration (log T_{90})  &  \sigma  (log T_{90}) &  w \\
            \noalign{\smallskip}
            \hline
            \noalign{\smallskip}
        short      &   -0.11   &  0.61   &   0.32   \\
        long      &     1.54   &  0.43   &   0.68     \\
            \noalign{\smallskip}
            \hline
         \end{array}
     $$ 
\begin{list}{}{}
\item[$^{\mathrm{ Average duration (log T_{90})
standard deviation (\sigma ) and weight (w) of the groups.
}}$] 
\end{list}
   \end{table}

A fit to the duration distribution has been taken using a maximum likelihood method
 with the superposition of two 
log-normal distributions.
This can be done by a standard search for 
5 parameters with $N=1929$ measured points (cf. \cite{pre}; Chapt. 15). 
Both log-normal distributions have two parameters; the fifth parameter 
defines the weight ($w_1$) of the first log-normal distribution. The second
weight is $w_2=(1-w_1)$ due to the normalization.
Therefore
we obtain the best fit to the 5 parameters through a maximum likelihood
 estimation (e.g., \cite{Ken}).
We search for the maximum of the formula

\begin{equation}
L = \sum_{i=1}^{N} \ln  \left( w_1 f_1(x_i,T_{1},\sigma_{1}) + 
w_2 f_2(x_i,T_{2},\sigma_{2} ) \right) 
\end{equation}

where
\begin{equation}
$$f_k = \frac{1}{ \sigma_k  \sqrt{2 \pi  }} 
\exp\left( - \frac{(x-T_k)^2}{2\sigma_k^2} \right) $$
\end{equation}

where $T_k$ is the mean in log $T_{90}$ and $\sigma $ is the standard deviation.
This fit gives us the
best parameters of the two-Gaussian fit (Table 1.),
which are very similar to previously published values
(\cite{HO}).

Secondly, a three-Gaussian fit has been taken with three $f_k$ functions
with eight parameters (three means, three standard deviations and two weights).
For the best fitted parameters see Table 2.
The best logarithm of the likelihood ($L_3$) is 12326.25 (\cite{Ken}).     
For two Gaussians the maximum of the likelihood was $L_2$=12320.11.
According to the mathematical theory, twice the difference of these 
numbers follows the $\chi  ^2$ distribution with three degrees of
freedom because the new fit has three more parameters

\begin{equation}
2 ( L_{3} -  L_{2}) \simeq \chi^2_{3},
\label{eq:chi}
\end{equation}

The difference is 6.14 which gives us a   0.5\%
probability.
Therefore the three-Gaussian fit is better and
 there is a 0.005 probability that it is caused by
statistical fluctuation.

   \begin{table}
 \caption[]{Three Gaussian fit of 
the GRBs duration distribution}
         \label{KapSou}
     $$ 
         \begin{array}{cccc}
            \hline
            \noalign{\smallskip}
     & Duration (log T_{90})  & \sigma  (log T_{90}) &  w \\
            \noalign{\smallskip}
            \hline
            \noalign{\smallskip}
          short   &  -0.25  &  0.53   &  0.26   \\
           long   &   1.55   & 0.42   &  0.68    \\
    intermediate  &   0.63   & 0.20   &  0.06      \\
            \noalign{\smallskip}
            \hline
         \end{array}
     $$ 
\begin{list}{}{}
\item[$^{\mathrm{ Average duration (log T_{90})
standard deviation (\sigma ) and weight (w) of the groups.
}}$] 
\end{list}
   \end{table}


\section{MONTE-CARLO SIMULATION}

One can check this 0.005 probability using the  Monte-Carlo (MC) simulation.
Take the two-Gaussian distribution with the best fitted parameters of the
observed data, and generate 1929 numbers for T$_{90}$
whose distribution follow the two-Gaussian distribution.
Find the best likelihood with five free parameters
(two means, two dispersions and two weights; but the sum of the last
two ones must be 1929). Secondly make a fit with three-Gaussian distribution
(eight free parameters, three means, dispersions and weights).
Take a difference between the two logarithms of the
maximum likelihood, which gives one number.

This procedure is
repeated 99 times and we have a hundred MC simulated
numbers. Only one of these numbers is bigger than
the value obtained from the BATSE 
 data (6.14). The distribution of
these differences are given in  Figure 2.
Therefore the MC simulations confirm the likelihood
law statement and gives us a similar probability (1\%) that
the third group is a statistical fluctuation.

   \begin{figure}
   \centering
\resizebox{\hsize}{!}
{\includegraphics[angle=-90,width=7cm]{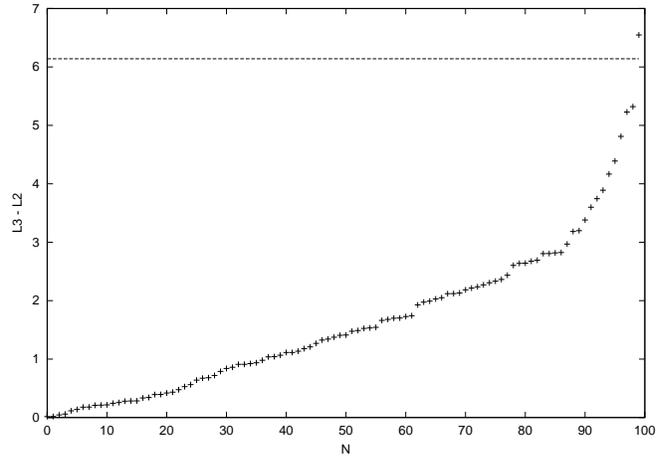}}
      \caption{Distribution of the MC simulated $L_3-L_2$.
$L_3$ is a likelihood with three-Gaussian and $L_2$
is a likelihood with two-Gaussian. }
   \end{figure}

Also the population of the third group generated by MC simulations 
is far from the GRB third group population which is 6\% (see Table 2.).
The average of the fluctuation population is 2,5\% and only one
of the hundred number is bigger than 6\%.

\section{ SYSTEMATICS}
In this section the possibility that the third intermediate
subgroup is an instrumental effect is discussed.
The BATSE on-board software tests for the existence of bursts by comparing 
the count rates to the threshold levels for  three separate time intervals: 
64 ms, 256 ms, and 1024 ms. The efficiency changes in the region of the middle 
area because the 1024 ms trigger is becoming less sensitive as burst 
durations fall below about one second. This means that at the ``intermediate"
timescale a large systematic deviation is possible. 
To reduce the effects of trigger systematics in this region we 
truncated the dataset to include only GRBs that would have triggered BATSE 
on the 64 ms timescale.

Using the Current BATSE catalog CmaxCmin table (\cite{M}) we choose the GRBs,
which numbers larger than one in the second column (64 ms scale
maximum counts divided by the threshold count rate).
This process reduced the bursts numbers very much,
therefore 
unfortunately just 958 bursts satisfied the above condition.

We repeated the maximum likelihood test with these
burst's durations. The computed probability is still
below 1\%. Therefore after eliminating some systematic
effects, the third group is still statistically significant.

\section{CONCLUSIONS}

   \begin{enumerate}
      \item 
It is possible that the three log-normal fit is accidental, and that there 
are only two types of GRBs. However, if the T$_{90}$ distribution of 
these two types of GRBs is log-normal, then the probability that the third 
group of GRBs is an accidental fluctuation is less than 0.5 \%.

      \item 
Therefore, statistically the third component existence is not questionable.
However, the physical existence of the third group is still argued.
The sky distribution of the third component is anisotropic as proven
by \cite{bal98}, \cite{bal99}, \cite{me00b}, \cite{li01}.
The logN-logS distribution is may also differ from the other group distribution
\cite{HO}. Opponently \cite{HA} believe the third statistically
proved subgroup is only a deviation caused by a complicated instrumental effect,
which can reduce some faint long burst's duration.
This paper does not deal with the effect mentioned above, 
however the triggering systematic effects are examined 
and after that the third group is still statistically significant.

      \item 
Therefore, this theme should be discussed in future papers
to further elucidate the reality and properties of the third class.

   \end{enumerate}

\begin{acknowledgements}
This research was supported in part through NATO advanced research
fellowship 1037/NATO/01, OTKA F029461, 
OTKA T034549. Useful discussions with L. G. Bal\'azs, J. T. Bonnell,
E. Fenimore, J. Hakkila, A.  M\'esz\'aros, P. M\'esz\'aros,  
are appreciated.
The author also thanks B. McBreen for useful
comments that improved the paper.

\end{acknowledgements}


\begin{thebibliography}{}


\bibitem[Balastegui et al. 2001]{br}
Balastegui, A., Ruiz-Lapuente, P. \& Canal, R. 2001, MNRAS, 328, 283

\bibitem[Bal\'azs et al. 1998]{bal98}
Bal\'azs, L.G., M\'esz\'aros, A. \& Horv\'ath, I. 1998, A\&A, 339, 1

\bibitem[Bal\'azs et al. 1999]{bal99}
Bal\'azs, L.G., M\'esz\'aros, A., Horv\'ath, I. \& Vavrek, R. 1999, 
A\&A Suppl., 138, 417

\bibitem[Hakkila, et al. 2000a]{HAK1} Hakkila, J., et al. 2000a,
Gamma-Ray Burst Fith Huntsville Symposium. Huntsville, Alabama. 
ed by R. M. Kippen, R. S. Mallozzi, and G. J. Fishman. AIP 526. 

\bibitem[Hakkila, et al. 2000b]{HAK2} Hakkila, J., et al. 2000b,
Gamma-Ray Burst Fith Huntsville Symposium. Huntsville, Alabama. 
ed by R. M. Kippen, R. S. Mallozzi, and G. J. Fishman. AIP 526. 

\bibitem[Hakkila, et al. 2000c]{HA} Hakkila, J., et al. 2000c,
ApJ, 538, 165

%
\bibitem[Horv\'ath, 1998]{HO} Horv\'ath, I. 1998, ApJ, 508, 757.

\bibitem[Kendall \& Stuart 1976]{Ken}
Kendall, M. \& Stuart, A. 1976, The Advanced Theory
of Statistics (Griffin, London)

\bibitem[Koshut, et al. 1996]{Kos} Koshut, T.M., et al. 1996, 
ApJ, 463, 570

\bibitem[Kouveliotou, et al. 1993]{K1} Kouveliotou, C., et al. 1993, 
ApJ, 413, L101
%
\bibitem[Kouveliotou, et al. 1995]{K} Kouveliotou, C., et al. 1995, 
Third Huntsville Symposium on GRB. New York: in AIP Conference Proceedings 
384, 84-89.

\bibitem[Litvin et al. 2001]{li01} Litvin, V.F., Matveev, S.A.,
Mamedov, S.V. \& Orlov, V.V. 2001,
Pis'ma v Astronomicheskiy Zhurnal, 27, 495 

\bibitem[McBreen, et al. 1994]{mcb} McBreen, B., Hurley, K.J.,
Long, R. \& Metcalfe, L. 1994, MNRAS, 271, 662

\bibitem[Meegan, et al. 2001]{M}  Meegan C. A., et al. 2001, Current BATSE 
Gamma-Ray Burst Catalog, on the Internet
http://www.batse.msfc.nasa.gov/data/grb/catalog/

\bibitem[M\'esz\'aros et al. 2000]{me00b}
M\'esz\'aros, A., Bagoly, Z., Horv\'ath, I., Bal\'azs, L.G. \&
Vavrek, R. 2000, ApJ, 539, 98


\bibitem[Mukherjee, et al. 1998]{MUK} Mukherjee, S., Feigelson, E.D., Babu, 
G.J., Murtagh, F., Fraley, C. \& Raftery, A. 1998, ApJ, 508, 314

\bibitem[Press et al. 1992]{pre} Press, W.H.,
Flannery, B.P., Teukolsky, S.A. \& Vetterling, W.T. 1992, 
Numerical Recipes (Cambridge University Press, Cambridge)


\end{thebibliography}
\end{document}